\documentclass{emulateapj}
\slugcomment{{\sc Accepted to ApJ:} December 31, 2014}
\usepackage{graphicx}
\usepackage{color}

\shorttitle{The origin of the red BSS sequence in M30}
\shortauthors{Xin et al.}

\begin{document}

\title{The binary mass transfer origin of the red blue straggler sequence in M30}

\author{Y. Xin\altaffilmark{1\dag}, F.R. Ferraro\altaffilmark{2},
  P. Lu\altaffilmark{1}, L. Deng\altaffilmark{1},
  B. Lanzoni\altaffilmark{2}, E. Dalessandro\altaffilmark{2},
  G. Beccari\altaffilmark{3}}
\affil{1. Key Laboratory of Optical Astronomy, National Astronomical
  Observatories, Chinese Academy of Sciences, Beijing, 100012, China}
\affil{2. Dipartimento di Fisica e Astronomia, Universit\'a degli
  Studi di Bologna, viale Berti Pichat 6/2, I-40127 Bologna, Italy}
\affil{3. European Southern Observatory, Karl Schwarzschild Strasse 2,
  D-85748 Garching bei Munchen, Germany}
\altaffiltext{\dag}{email: xinyu@nao.cas.cn. The work is supported by
  the National Natural Science Foundation of China through grant
  Y111221001 and the 973 Program 2014CB845702.}

\begin{abstract}
Two separated sequences of blue straggler stars (BSSs) have been
revealed by Ferraro et al. (2009) in the color-magnitude diagram (CMD)
of the Milky Way globular cluster M30.  Their presence has been
suggested to be related to the two BSS formation channels (namely,
collisions and mass-transfer in close binaries) operating within the
same stellar system.  The blue sequence was indeed found to be well
reproduced by collisional BSS models. In contrast, no specific models for
mass transfer BSSs were available for an old stellar system like
M30. Here we present binary evolution models, including case-B mass
transfer and binary merging, specifically calculated for this cluster.
We discuss in detail the evolutionary track of a $0.9+0.5 M_\odot$
binary, which spends approximately 4 Gyr in the BSS region of the CMD
of a 13 Gyr old cluster. We also run Monte-Carlo simulations to study
the distribution of mass transfer BSSs in the CMD and to compare it
with the observational data. Our results show that: (1) the color
and magnitude distribution of synthetic mass transfer BSSs defines 
a strip in the CMD that nicely matches the observed red BSS
sequence, thus providing strong support to the mass transfer origin
for these stars; (2) the CMD distribution of synthetic BSSs never
attains the observed location of the blue BSS sequence, thus
reinforcing the hypothesis that the latter formed through a different
channel (likely collisions); (3) most ($\sim 60\%$) of the synthetic
BSSs are produced by mass-transfer models, while the remaining $<
40\%$ requires the contribution from merger models.
\end{abstract}


\section{Introduction}
Blue straggler stars (BSSs) are commonly defined as stars brighter and
bluer than the main-sequence (MS) turnoff in the host stellar cluster.
They are thought to be central H-burning stars, more massive
than the MS turnoff stars (Shara et al. 1997; Gilliland et al. 1998; 
De Marco et al. 2004; Fiorentino et al. 2014).  In stellar
  systems with no evidence of recent star formation, their origin
  cannot be explained in the framework of normal single-star evolution
  and two main formation channels are currently favored: (1) mass
  transfer (MT) in binary systems, possibly up to the complete
  coalescence of the two stars, and (2) stellar collisions.
Both these processes can potentially bring new hydrogen into the core
and therefore ``rejuvenate'' a star to its MS stage (e.g., Lombardi et
al. 1995, 2002; Chen \& Han 2009).

The scenario of physical collision was originally presented by Hills
\& Day (1976), and then different colliding encounters including
single-single (Benz \& Hills 1987), single-binary, and binary-binary
processes (Leonard 1989; Ouelette \& Pritchet 1998) have been
investigated in subsequent work. Collisions are believed to be
important especially in dense environments, such as the cores of
globular clusters (GCs; Bailyn 1992; Ferraro et al 1995, 1997,
2003a,b) and even the center of some open clusters (Leonard \&
Linnell 1992; Glebbeek et al. 2008).

McCrea (1964) first proposed that BSSs can be formed through MT in
close binaries. The process starts when the primary fills up its Roche
Lobe and then transfers material to the secondary through the inner
Lagrangian point.  The secondary increases its mass and shows up as a
BSS when its luminosity exceeds that of the MS turnoff stars.  There
are three cases of MT, defined according to the evolutionary stage of
the primary when it starts to transfer mass to the secondary
(Kippenhahn \& Weigert 1968): case-A when the primary is on the MS,
case-B when it is in post-MS but before helium ignition, and case-C
during central He-burning and thereafter.  Tian et al. (2006) reveal
that the BSSs in M67 can be effectively generated in short-period
binaries via case-A and case-B MT. Lu et al. (2011) pay close
attention to BSS populations in the intermediate-age star clusters,
where they find that both case-A and case-B MT can produce BSSs, and
BSSs via case-B are generally bluer and even brighter than those from
case-A.  MT in binaries might be the dominant formation channels in
all environments (e.g., Knigge et al. 2009; Leigh et al. 2013), and
most likely it is so in low-density GCs, open clusters and the
Galactic field (Ferraro et al. 2006a; Sollima et al. 2008; Mathieu et
al. 2009; Preston \& Sneden 2000). In the case of case-B or case-C binary origin, 
a BSS with a white dwarf companion is expected. This has been recently
confirmed for three objects in the open cluster NGC 188, thanks to HST
ultraviolet observations (Gosnell et al. 2014). MT can also produce
anomalous surface composition on the accretor's surface.  Indeed C and
O depletion has been observed in the atmosphere of a sub-sample of
BSSs in the GCs 47 Tucanae, M30 and $\omega$ Centauri (Ferraro et
al. 2006b; Lovisi et al. 2013; Mucciarelli et al. 2014), thus likely
indicating the MT origin of these stars.

Because of their large number of member stars, GCs are the ideal
environment for BSS studies. Nominally all the GCs observed so far
have been found to harbor a significant number of BSSs (Piotto et
al. 2004; Leigh et al. 2007).  Moreover, the stellar density in GCs
varies dramatically from the central regions to the outskirts, and
since BSSs in different environments (low- versus high-density) could
have different origins (e.g., Fusi Pecci et al. 1992; Davies et
al. 2004), these stellar systems allow to investigate both formation
channels simultaneously. However a clear distinction is hampered by
the internal dynamical evolution of the parent cluster (Ferraro et
al. 2012). In fact, having masses larger than normal cluster stars,
BSSs are affected by dynamical friction, a process that drives the
objects more massive than the average toward the cluster center, over
a timescale which primarily depends on the local mass density (e.g.,
Alessandrini et al. 2014).  Hence, as the time goes on, heavy objects
(like BSSs) orbiting at larger and larger distances from the cluster
center are expected to drift toward the core: as a consequence, the
radial distribution of BSSs develops a central peak and a dip, and the
region devoid of these stars progressively propagates outward.

Ferraro et al. (2012) used this argument to define the so-called
``dynamical clock'', an empirical tool able to measure the dynamical
age of a stellar system from the shape of its BSS radial
distribution. This appears indeed to provide a coherent interpretation
of the variety of BSS radial distributions observed so far: GCs with a
flat BSS radial distribution (Ferraro et al. 2006b; Dalessandro et
al. 2008a; Beccari et al. 2011) are dynamically young systems, GCs
with bimodal distributions (e.g., Ferraro et al. 1993, 2004; Lanzoni et
al. 2007a; Dalessandro et al. 2008b; Beccari et al. 2013, and
references therein) are dynamically intermediate-age systems (their
actual dynamical age being determined by the distance of the dip of the
distribution from the cluster center), and GCs with a single-peaked
BSS distribution (Ferraro et al. 1999a; Lanzoni et al. 2007b;
Contreras~Ramos et al. 2012; Dalessandro et al. 2013) are dynamically
old systems.  Following this view, in the cluster center we expect a
mixed BSS population: collisional BSSs produced by stellar
interactions and MT BSSs drifted to the center because of dynamical
friction.

A recent discovery has potentially opened the possibility to
photometrically distinguish collisional BSSs from MT BSSs in the same
cluster.  Two well distinct BSS sequences, almost parallel and
similarly populated, have been found in the color-magnitude diagram
(CMD) of M30 (Ferraro et al. 2009, hereafter F09).  Similar
  features have been detected also in NGC~362 by Dalessandro et
  al. (2013) and NGC~1261 by Simunovic et al. (2014).
The blue-BSS sequence
observed in M30 is nicely reproduced by collisional isochrones (Sills
et al. 2009) with ages of 1-2 Gyr.  In contrast, the red-BSS
population is far too red to be consistent with collisional isochrones
of any age, while it seems to well correspond to the ``low-luminosity
boundary'' outlined by the MT binary populations simulated by Tian et
al. (2006) for the open cluster M67 (with an age of 4 Gyr), once
``extrapolated'' to the case of a much older ($\sim 13$ Gyr) cluster
as M30. However, given the large difference in age between these two
systems (which implies significantly different mass regimes for their
current binary systems and BSSs), such an extrapolation can be risky
and it was made only because no specific models of MT binaries were
available at that time for the case of M30.

With the aim of finally providing a firm conclusion about the true
nature of the red-BSS sequence observed in M30, here we present MT
binary models specifically calculated for this cluster. We also used
Monte-Carlo simulations to study the distribution of synthetic MT-BSSs
in the CMD. The details of the binary evolution models are described
in Sect. 2.  The results of the Monte-Carlo simulations are presented
and compared with the observations in Sect. 3.  Sect. 4 gives summary
and discussion.

\section{The model of primordial binaries}
We use the stellar evolution code originally written by Eggleton
(1971, 1972, 1973) and then updated several times (e.g., Han et
al. 1994; Pols et al. 1995, 1998) to calculate the evolution of
primordial binaries. The detailed description of the version we used
can be found in Han et al. (2000) and Lu et al. (2010). In
  particular, for this work we adopted the radiative opacity of
Iglesias \& Rogers (1996) and the molecular opacities of Alexander \&
Ferguson (1994), and conservations in both mass and angular momentum
are assumed.

Roche lobe overflow (RLOF) is used as a boundary condition in the
code. When RLOF takes place, the mass transfer rate ($dm/dt$) between
the two components is described as:
\begin{equation}
\footnotesize
\frac{dm}{dt}=const. \times {\rm max}[0,~(R_{\rm star}/R_{\rm lobe} -1)^3],
\end{equation}
where $R_{\rm star}$ is the radius of the donor (primary star),
and $R_{\rm lobe}$ is the effective radius of the corresponding Roche
lobe.  We use $const. = 500 M_\odot$ yr$^{-1}$ to keep the RLOF
steady. The donor may overfill its Roche lobe, but the condition
  of $(R_{\rm star}/R_{\rm lobe} - 1) \leq 0.001$ must be satisfied
(Han et al. 2000).  In order to avoid numerical instabilities,
the calculation stops if the RLOF is unstable.

We assume the initial orbital eccentricity to be zero ($e=0$) for all
the models. Convective overshooting is not considered as it can barely
influence the evolution of low-mass stars (Pols et al.  1998).  After
all the fundamental ingredients are prepared, instead of simply
approximating the major parameters at each time step by interpolating
between corresponding evolutionary tracks, we performed the
calculation in a more precise way, exactly following the evolution of
both components.  Eggleton's code provides details for both components
of a binary, but it loses the information of the secondary during the
RLOF.  We made minor modifications to the code, so that the mass-loss
history of the donor is recorded, and it is used to compute the
subsequent evolution of the accretor.  Such a treatment keeps the
evolution of both components synchronized. The accreted material from
the donor is assumed to be deposited onto the surface of the accretor
and instantly distributed homogeneously over its outer layer.

In this work we initially considered both case-A and case-B MT.
However we found that most of the case-A MT binaries did not survive
until the current cluster age. Moreover, the MT was often unstable,
thus making the code stop and preventing us to follow the complete
evolutionary paths of these systems. Hence, in the following we focus
only on case-B MT (and binary mergers).  In any case, BSSs
generated by case-A MT have been found to lie on the red side of the
case-B locus (Lu et al.  2011), and they are therefore expected to not
affect the main result of this work.  Given the age and the
metallicity of M30 ($\sim$13~Gyr and Z=0.0001), its current turnoff
mass is $\sim0.75~M_\sun$. Hence, the currently observable BSSs
generated from case-B MT in primordial binaries should come from donor
stars with initial masses roughly between 0.7 and 1.1~$M_\sun$.  We
can trace case-B MT only if the donor is evolving along the sub-giant
branch when the RLOF occurs. The code stops if it is on the red giant
branch phase due to numerical instability. It also stops if the sum of
the two components' radius is equal to or larger than the orbital
radius, which marks the moment when a merger occurs.  As the
modifications of internal structure and chemical profile are quite
complicated during a merging process and still unclear in knowledge,
 we consider the merger product as a single star with mass equal
  to the sum of the two components' masses and evolved from its zero
  age main sequence (ZAMS) to an age equal to 13 Gyr minus the time of
  the merger.  The hydrogen fuel in its core is calculated as the sum
  of the central hydrogen left in the two components when merging
  happens.

As an illustrative example of a binary system experiencing MT and
producing a BSS in M30, we calculated a binary with metallicity
appropriate for the cluster ($Z=0.0001$; Ferraro et al. 1999b),
composed of a primary (donor) with $0.9 M_\sun$ and a secondary
(accretor) with $0.5 M_\sun$, and having an initial orbital radius of
$2.7~R_\sun$. Figure 1 shows the evolutionary tracks of the two
components (dashed and dotted lines for the primary and the secondary,
respectively). Both tracks are interpolated using the time step of the
component for which the calculation stops first. We also report a few
symbols along the evolutionary tracks to flag reference events in the
evolution of the two components: (1) the beginning of the MT process
(open squares), (2) the epoch at which the mass ratio ($q$) is equal
to one (open stars), (3) the end of the MT process (solid squares),
and (4) the location at 13 Gyr (open circles). It is shown that the
two components severely depart from the regular evolutionary tracks
after MT begins.  Actually, according to case-B MT, the
primary is just leaving its MS when the RLOF occurs, at the age of
$\sim 7.54$ Gyr (open square along the dashed line).  After MT stops
at the age of $\sim 12.78$ Gyr (filled square along the dashed line),
the primary follows the evolutionary behavior of a single star again
and eventually evolves into a white dwarf. In the meanwhile, the
secondary gains mass and becomes progressively more luminous.  MT
stops while the secondary is still in its MS phase (solid square in
the dotted line). Both tracks in Figure 1 are truncated at about
  1 Gyr after the end of the MT process. As the binary system starts
  MT at 7.54 Gyr and M30 has an age of 13 Gyr, the corresponding
  theoretical isochrones are also plotted for reference as solid lines
  in the figure. Table 1 presents the main parameters of the two
components at five key epochs during the binary's evolution.

\begin{figure}
\centering
\includegraphics[width=8cm]{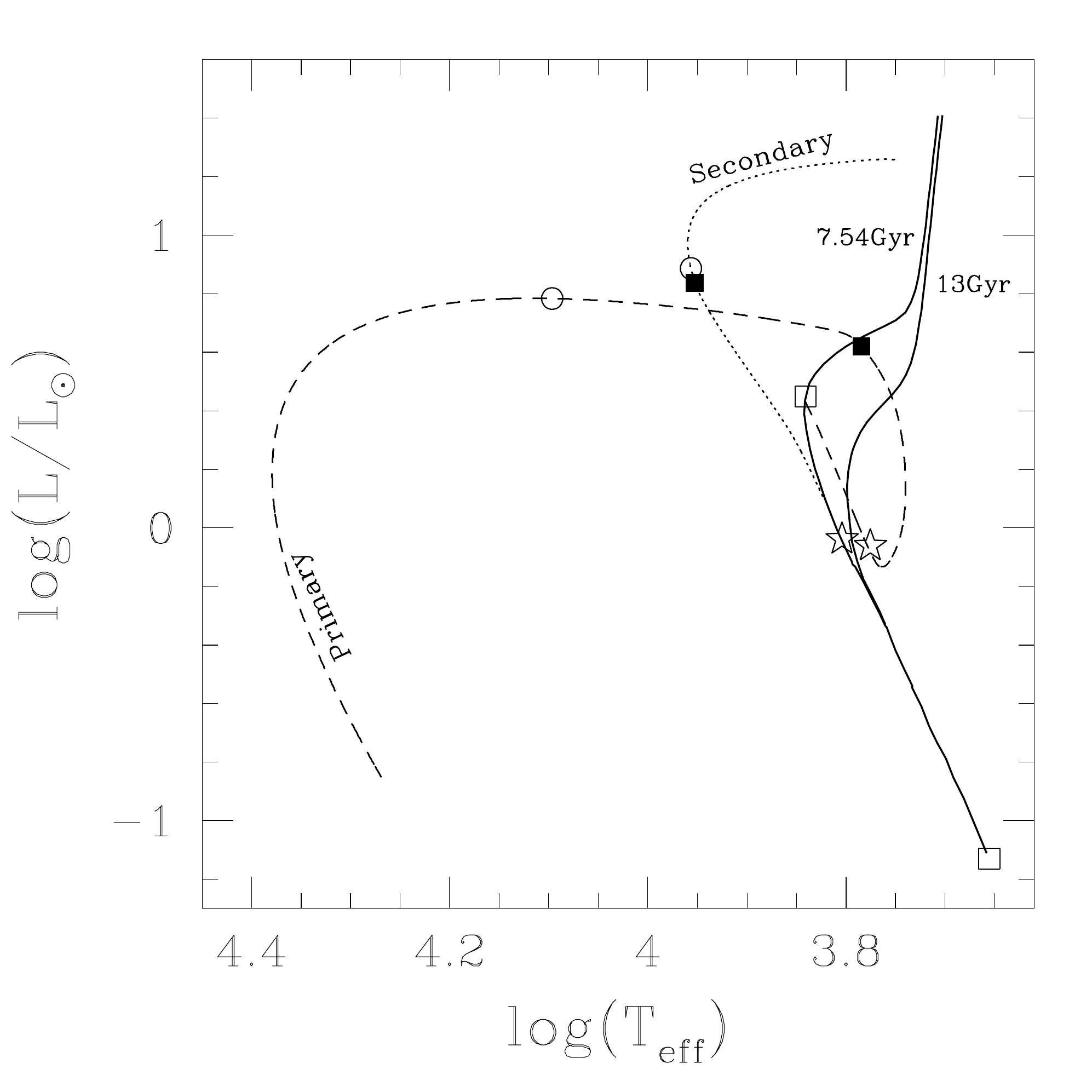}
\caption{Evolutionary tracks for an illustrative example of binary
  system ($0.9 M_\sun +0.5 M_\sun$, initial orbital radius
  $=2.7~R_\sun$) undergoing case-B mass transfer. The evolutionary
  tracks of the primary (donor) and secondary (accretor) stars are
  shown as dashed and dotted lines, respectively. Different symbols
  mark a few key events in the binary evolution: the beginning of MT
  (open squares), the epoch at which the mass ratio is equal to one
  (open stars), the end of the MT process (solid squares), and the 13
  Gyr age (open circles).  For the sake of comparison, the 7.54 and 13
  Gyr isochrones at the appropriate metallicity of M30 (Z=0.0001) are
  also plotted in solid lines. }
\end{figure}

\begin{table*}
\begin{center}
\footnotesize
\caption{Main parameters of the illustrative binary system ($0.9 M_\sun + 0.5 M_\sun$).}
\begin{tabular}{rccccccccc}
\tableline\tableline
Epoch & Age    & P   & a           & Mass       & $log (L/L_\sun)$ & $log (T_{\rm eff})$ & $X_C$ & $Y_C$ & \.M \\
      & (Gyr)  & (d) & ($R_\sun$)  & ($M_\sun$) &                 &             &       &       & ($M_\sun$ yr$^{-1}$) \\
\tableline
1 &0.0000  & 0.4345 & 2.7000 & 0.9000 & -0.0193 & 3.8108 & 0.7700 & 0.2299 & 0.0  \\
  &        &        &        & 0.5000 & -1.1571 & 3.6524 & 0.7700 & 0.2299 &      \\
2 &7.5412  & 0.4345 & 2.7000 & 0.9000 &  0.4826 & 3.8496 & 0.0485 & 0.9514 & $2.6746\times10^{-15}$ \\
  &        &        &        & 0.5000 & -1.1356 & 3.6544 & 0.7029 & 0.2970 &      \\
3 &7.5491  & 0.3366 & 2.2772 & 0.6999 &  -0.0613 & 3.7783 & 0.0481 & 0.9518 & $3.1974\times10^{-8}$ \\
  &        &        &        & 0.6999 & -0.0394 & 3.7937 & 0.7086 & 0.2913  & \\
4 & 12.7828& 1.9555 & 7.3598 & 0.2349 & 0.6213 & 3.7845 & 0.0000 & 0.9999 & 0.0 \\
    &      &        &        & 1.1663 & 0.8375 & 3.9526 & 0.0843 & 0.9157 & \\
5 &13.0000 & 1.9555 & 7.3598 & 0.2349 & 0.7841 & 4.0965 & 0.0000 & 0.9999 &  0.0 \\
  &        &        &        & 1.1663 & 0.8870 & 3.9564 & 0.0313 & 0.9687 & \\
\tableline
\end{tabular}
\tablecomments{The columns are: (1) Reference epoch number, (2) age in
  Gyr, (3) orbital period in days, (4) separation between the two
  components, (5) mass in $M_\sun$, (6) luminosity in
  $log~(L/L_\sun)$, (7) logarithm of the effective temperature, (8)
  hydrogen mass fraction in the core, (9) helium mass fraction in the
  core, (10) MT rate in $M_\sun$ yr$^{-1}$.  The epoch indicates: (1)
  the ZAMS, (2) the beginning of MT, (3) the time when the mass ratio
  is equal to 1, (4) the end of MT, (5) the time is equal to 13~Gyr. 
  Two lines in each epoch indicate the parameters of the primary
  (up) and the secondary (low) components, respectively.}
\end{center}
\end{table*}

\begin{figure}
\centering
\includegraphics[width=8cm]{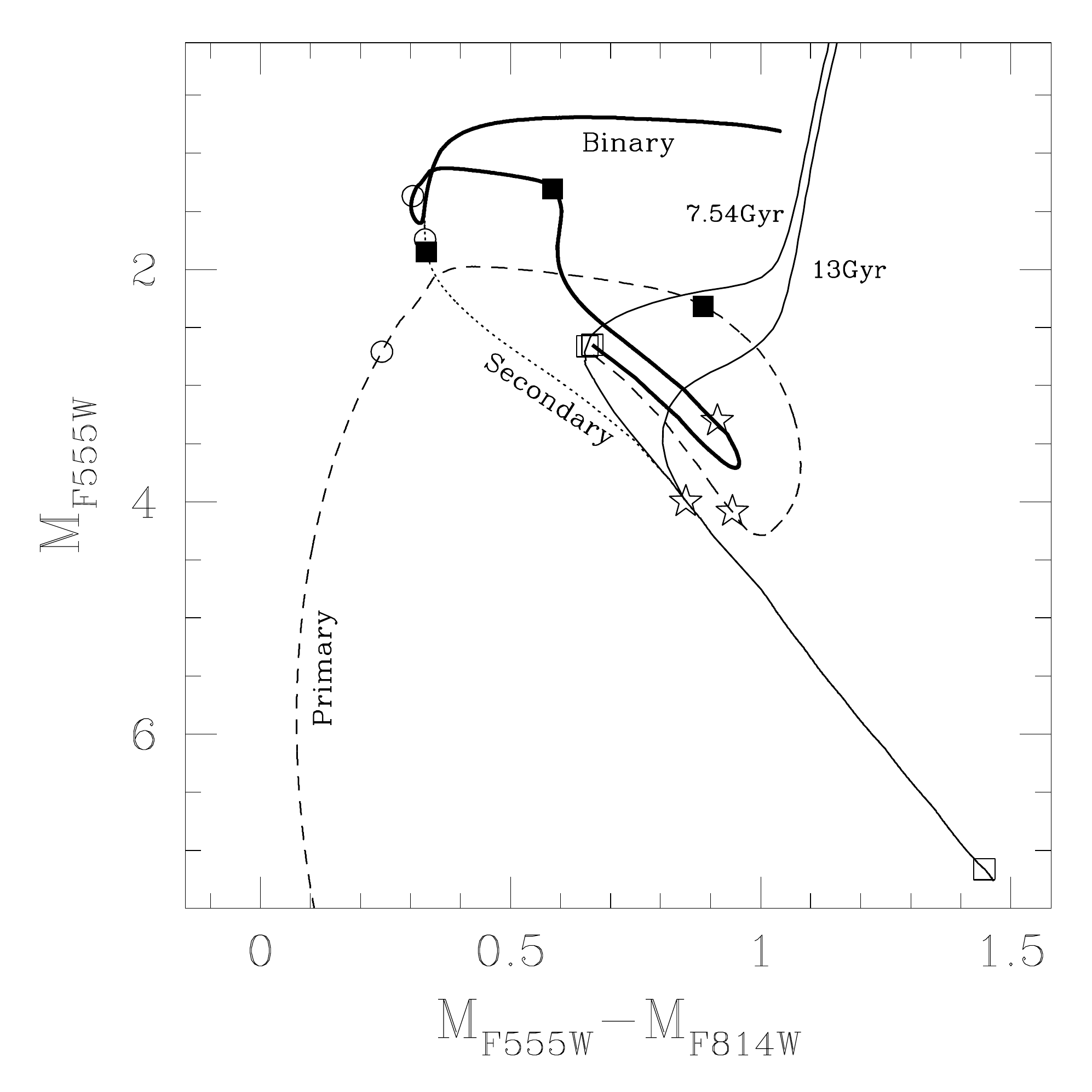}
\caption{As in Figure 1, but in the absolute CMD and with added the
  track of the binary system (thick solid line), obtained by combining
  the luminosity contributions of the primary and secondary
  components. The two thin solid lines are the 7.54 and 13~Gyr
    isochrones.}
\end{figure}

\begin{figure}
\centering
\includegraphics[width=8cm]{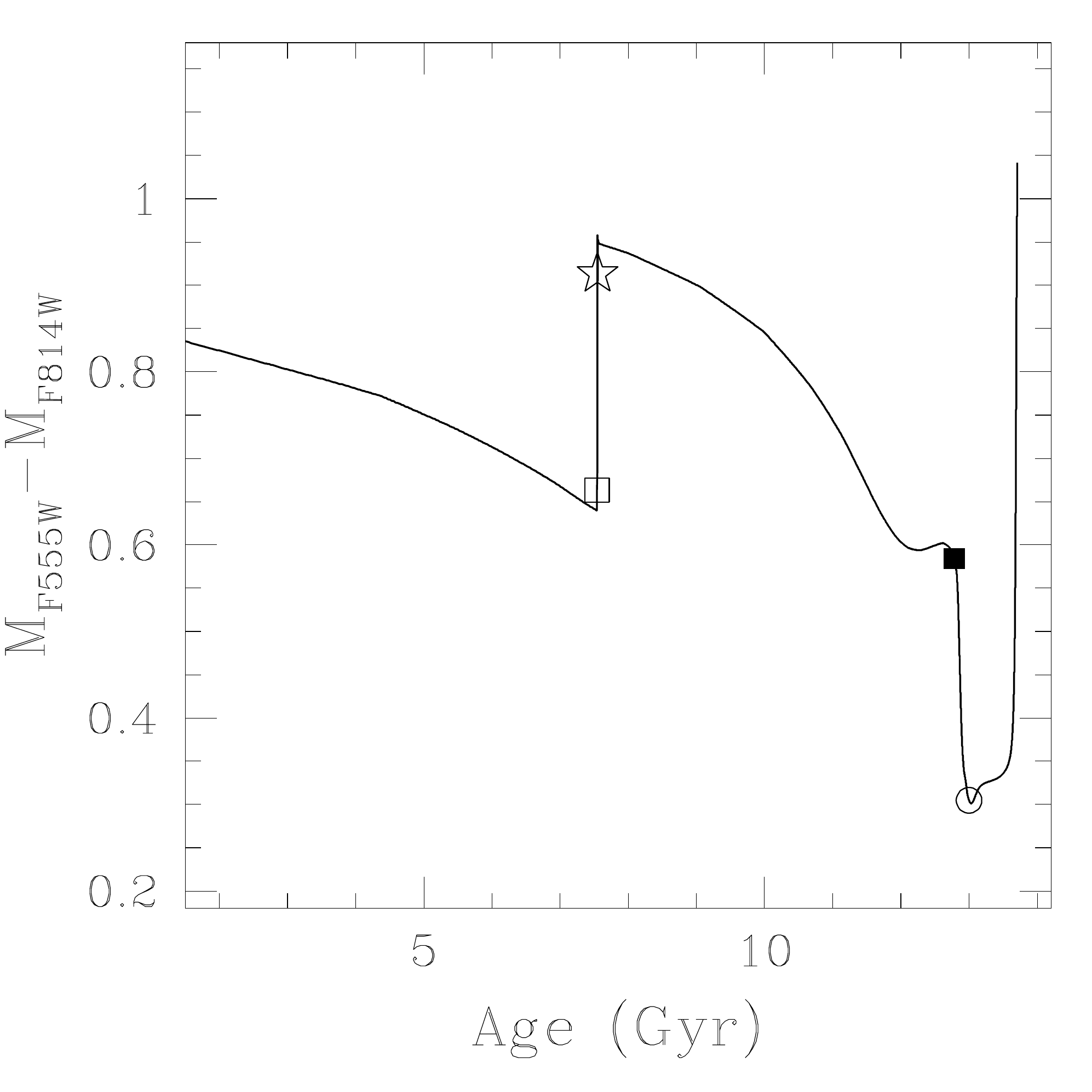}
  \caption{Time evolution of the color of the synthetic binary system
  considered in Figures 1 and 2. All the symbols keep the same meaning
  as in the previous figures. }
\end{figure}

Of course the photometric properties of the individual components of a
binary system cannot be distinguished at the distance of Galactic GCs,
and only the combined luminosity of the two components is
observed. Hence, in order to perform an appropriate comparison with
the observations, for each synthetic BSS we computed the total light
as the sum of the luminosities of the two components, at each
evolutionary step.  To this end, we used our set of models, which
describe the properties (e.g., effective temperature, luminosity, mass and
surface gravity) of each component of the binary system along their
evolutionary tracks. In particular, from the effective temperature and
gravity we constructed synthetic spectrum for each component by
linearly interpolating between relevant spectra from the $\tt
{ATLAS9}$ library (Castelli \& Kurucz 2004). Then, the magnitudes have
been obtained by convolving the spectrum with the throughput curves of
the photometric filters.  Specifically for this work, as the
observational data for M30 have been acquired in the F555W and F814W
filters of the WFPC2 on board the HST, the throughput curves have been
directly downloaded from the $\tt {STScI}$ website\footnote{\small
  http://www.stsci.edu}.  The zero-points have been calculated by
making the $V$-band magnitude of Vega equal to 0.03, and all its
colors equal to 0.00.  Also the Vega spectrum has been extracted from
the $\tt {ATLAS9}$ library.  The synthetic magnitude of two components
is calculated with the following formula:
\begin{equation}
\footnotesize
M_i = M_{i,1} - 2.5 \times \log(1+10^{\frac{M_{i,1} - M_{i,2}}{2.5}}),
\end{equation}
where $M_i, M_{i,1}$, and $M_{i,2}$ are the $i$-band magnitudes of the
binary, the primary, and the secondary, respectively.
 
The evolutionary track of the selected binary system in the absolute CMD
(M$_{\rm F555W}-$M$_{\rm F814W}$, M$_{\rm F555W}$) is presented in
Figure 2 in thick solid line.  As in Figure 1, the dashed and the
dotted lines are the tracks of the primary and the secondary,
respectively. The two thin solid lines are the 7.54 and 13 Gyr
  isochrones. The figure clearly shows that before the inversion of
the mass ratio (star symbols), the (higher luminosity) primary
component dominates the synthetic track of the binary system, but
after that moment, the secondary becomes dominating and the binary
system evolves towards the BSS region in the CMD. Figure 3 shows the
evolution of the color of the synthetic binary as a function of time.
The combined analysis of Figures 2 and 3 suggests that, after a
  short initial period during which the luminosity of the binary
  decreases because the (dominant) primary component is losing mass,
  the system starts to become progressively brighter and bluer with
  time. At an age of $\sim 10$ Gyr the binary is brighter and bluer
  than the MS turnoff point of the cluster population, thus finally
  appearing as a BSS. The system remains in the BSS region of the CMD
  for the following 3-4 Gyr, before deviating toward the red giant
  branch.

\section{Distribution of synthetic MT-BSSs in the CMD}
In order to construct a population of synthetic BSSs originated
  from MT binaries and compare its distribution in the CMD with the
  observations of M30, we adopted the following procedure:
\begin{itemize}
\item[$(i)$] we build a grid of binary evolution models formally
  including all possible binary systems able to generate a BSS
  currently observable in M30;
\item[$(ii)$] we randomly generated a large number of binaries by
  extracting the values of the three basic parameters characterizing
  each system (the masses of the two components and the orbital
  separation) from appropriate distribution functions;
\item[$(iii)$] for those binaries generated in step $(ii)$ and covered 
  by the grid constructed in step $(i)$, we identified 
  their most appropriate stellar models by multi-dimensional
  interpolation within the grid;
\item[$(iv)$] all such binaries (from step $(iii)$) that appear to be brighter and bluer
  than the MS turnoff of M30 have been retained as BSSs and included
  in what we call the ``synthetic BSS reservoir'', and that is used for the
  comparison with the observations.
\end{itemize}
The following Sections provides the details of such a procedure.

\subsection{Building up the grid of models}
In order to investigate how BSSs originated via MT in primordial
binaries populate the CMD of an old GC (as M30), a grid including all
the models of primordial binaries that have started MT and/or merging
process and can survive at least for 13~Gyr is required.  We therefore
calculated binary models assuming suitable combinations of their basic
parameters, each model (either ``pure'' MT, or binary merger)
representing a node of the grid.  As the work focuses on the
contribution of case-B MT to the BSS population in a star cluster like
M30, and the turnoff mass of a 13~Gyr isochrone is $\sim0.75 M_\sun$,
the range of initial mass for the primary is set to 0.7-1.1 $M_\sun$,
with steps of 0.1 $M_\sun$. The upper limit, $1.1 M_\sun$, is used
because a binary system with 1.2 $M_\sun$ primary can barely survive
13~Gyr.  The initial mass range of the secondary is wider: we use
0.3-1.1 $M_\sun$, with steps of 0.1 $M_\sun$. The lower limit is set
by the minimum mass of a secondary that, in 13~Gyr, can experience MT
from a primary in the adopted mass range.  The orbital separation
ranges from 1.0 to 10 $R_\sun$, with intervals of 0.1 $R_\sun$. The
ranges set for these three parameters cover all the possibilities of
making a BSS from a case-B MT binary in an old cluster as M30. Note
that in a few cases the code stops because the evolutionary stage is
numerically unstable. Thus a few nodes of the grid can be
missed. However the adopted interpolation procedure (see Section 3.3)
between nearby nodes fully recovers this problem.  Table 2 gives an
excerpt from the overall grid.  

\begin{table}
\begin{center}
\footnotesize
\caption{Excerpt from the grid of binary models} 
\begin{tabular}{rccccccccc}
\tableline\tableline
 age & $M_{1,g}$ & $M_{2,g}$ & $a_g$ & $M_{\rm F555W}$  & $M_{F55\rm 5W}-M_{\rm F814W}$ \\
(Gyr)  & ($M_\sun$) & ($M_\sun$) & ($R_\sun$)  &  (mag) &   (mag)       \\
\tableline
13.0 &  0.8  & 0.3  & 2.8 &  4.22 & 0.76 \\
13.0 &  0.8  & 0.3  & 2.9 &  4.14 & 0.74 \\
13.0 &  0.8  & 0.3  & 3.0 &  4.06 & 0.74\\
13.0 &  0.8  & 0.4  & 1.9 &  4.83 & 0.85 \\
13.0 &  0.8  & 0.4  & 2.0 &  4.73 & 0.83 \\
13.0 &  0.8  & 0.4  & 2.1 &  4.62 & 0.81 \\
13.0 &  0.8  & 0.4  & 2.2 &  4.51 & 0.80 \\
13.0 &  0.8  & 0.4  & 2.3 &  4.41 & 0.78 \\
13.0 &  0.9  & 0.5  &  2.0 & 3.53 & 0.63 \\
13.0 &  0.9  & 0.5  &  2.1  & 3.39 & 0.58 \\
\tableline
\end{tabular}
\tablecomments{The columns are: (1) current age, (2) initial mass of
  the primary, (3) initial mass of the secondary, (4) initial orbital
  separation of the two components, (5) HST/WFPC2 F555W magnitude of
  the binary at the indicated age, (6) HST/WFPC2 (F555W-F814W) color
  of the binary at the indicated age. } 
\tablecomments{Table 2 is published in its entirety both in the
    electronic edition of the Astrophysical Journal and at the web
    site http://www.cosmic-lab.eu/Cosmic-Lab/MT-BSS.html. A portion
    is shown here for guidance regarding its form and content.}

\end{center}
\end{table}

\subsection{Generating binary systems: distribution functions}
The basic parameters characterizing a binary system are the
  masses of the two components (or, equivalently, the total binary
  mass and the component mass ratio) and the orbital separation. In
  our procedure, each binary is generated by randomly extracting the
  initial values of these parameters from the following distribution
  functions. 
\begin{itemize}
\item[1)]{Assuming the initial mass function of Kroupa et al. (1991),
  we generate binaries with total mass $M$ (in units of $M_\odot$)
  extracted from the following function (see also Hurley et al. 2001):
\begin{equation}
\scriptsize
M(X)=0.33\times\left[{\frac{1}{(1-X)^{0.75}+0.04\times(1-X)^{0.25}}}-{\frac{(1-X)^2}{1.04}}\right]
\end{equation}
where $X$ is a random number uniformly distributed between 0 and 1.
The values of $M(X)$ are limited between 0.2 and $100 M_\sun$, based
on the assumption that for ``traditional'' single star populations,
the initial mass coverage ranges between 0.1 and $50 M_\sun$.}
\item[2)]{The mass ratio distribution for binaries in GCs is still
  quite controversial. Following Hurley et al. (2001), we adopt a
  uniform distribution between the following two limits:
\begin{equation}
\footnotesize
1>q>max\left[{\frac{0.1}{M(X)-0.1}}, 0.02 (M(X)-50.0) \right]
\end{equation}}
The masses of the two components are then obtained from the total
binary mass $M(X)$ and the value of $q$.
\item[3)]{For the orbital separation, the flat distribution in $\log
  a$ assumed by Pols \& Marinus (1994) is used here. The minimum size
  of $a$ corresponds to the value when a ZAMS star fills its Roche
  lobe. The maximum size is 50 AU.}
\end{itemize}

\subsection{Monte-Carlo simulations: comparison with the observations}
Monte-Carlo simulations were performed by randomly extracting
  $10^6$ values of the binary basic parameters from the distribution
  functions described in Section 3.2.  This provided us with $10^6$
  binary systems, each characterized by a group of ($M_1$, $M_2$ and
  $a$) values. 
  Then, the binaries covered by the grid were extracted, and their corresponding
  stellar evolutionary status at 13~Gyr were determined by means of a
  multi-dimensional interpolation among the nodes of the grid
  confining the values of initial $M_1, M_2$ and $a$. 
  Since some of the grid nodes are merger models (see Section
  3.1), we flag as ``merger-important'' those binaries for which at
  least half of the nodes used during the interpolation corresponds to
  mergers. 

\begin{figure*}
\centering
\includegraphics[width=10cm]{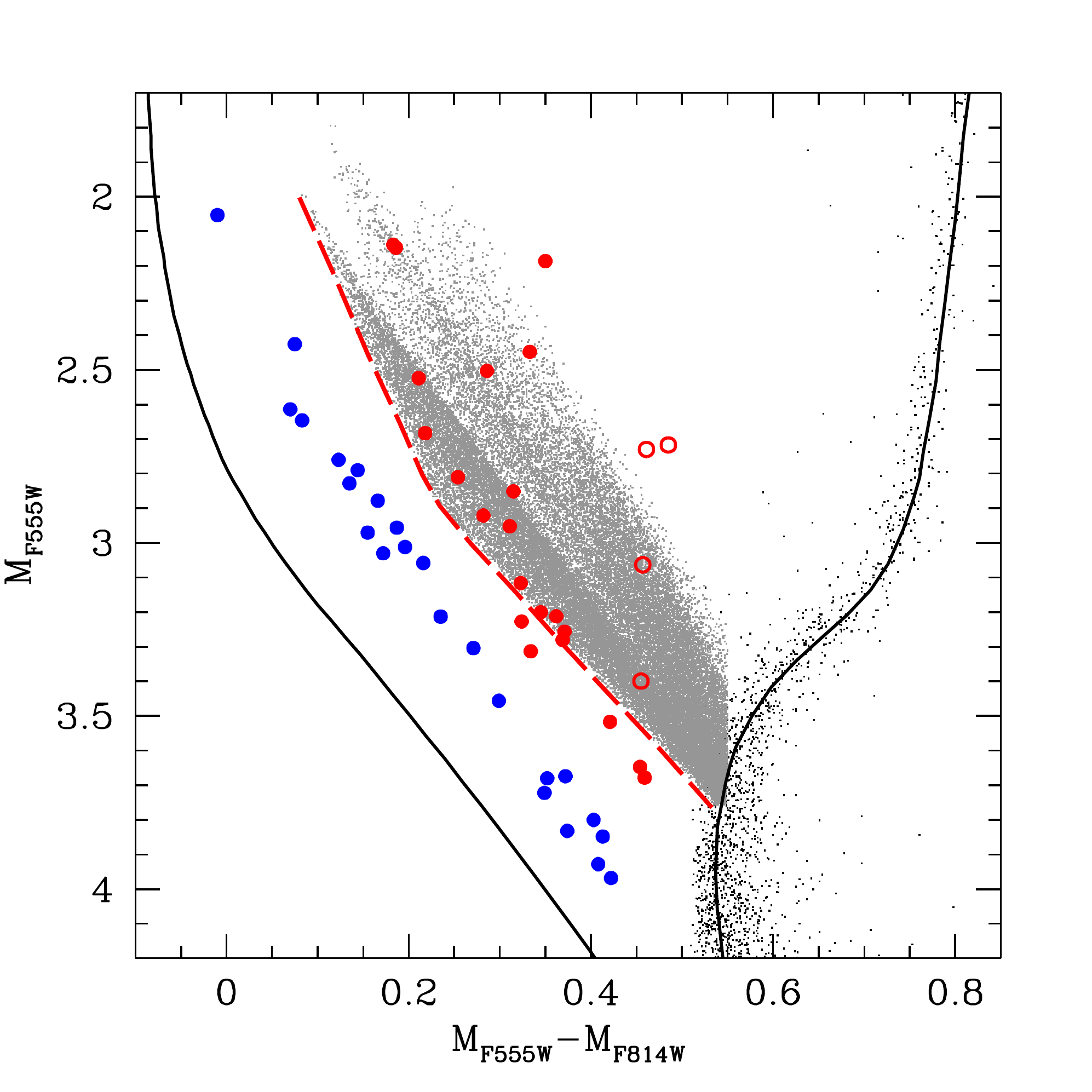}
\caption{Distribution of the synthetic
      MT-binaries (grey dots) bluer than the cluster MS turnoff, in
      the absolute CMD of M30. The red dashed line marks the
      ``low-luminosity boundary'' of the distribution. The BSSs
      observed by F09 along the red and the blue sequences are shown
      as solid red and blue circles, respectively. The positions of 4
      additional red-BSSs, not considered in F09, are also shown as
      empty red circles.  Clearly, the distribution of synthetic
      BSSs well samples the location of the observed red-BSSs, and its
      low-luminosity boundary nicely follows the red-BSS sequence. The
      black lines, shown for reference, are the 13~Gyr isochrone best
      fitting the observed M30 data (black dots) and the ZAMS from
      F09. }
\end{figure*}

\begin{figure*}
\centering
\includegraphics[width=11cm]{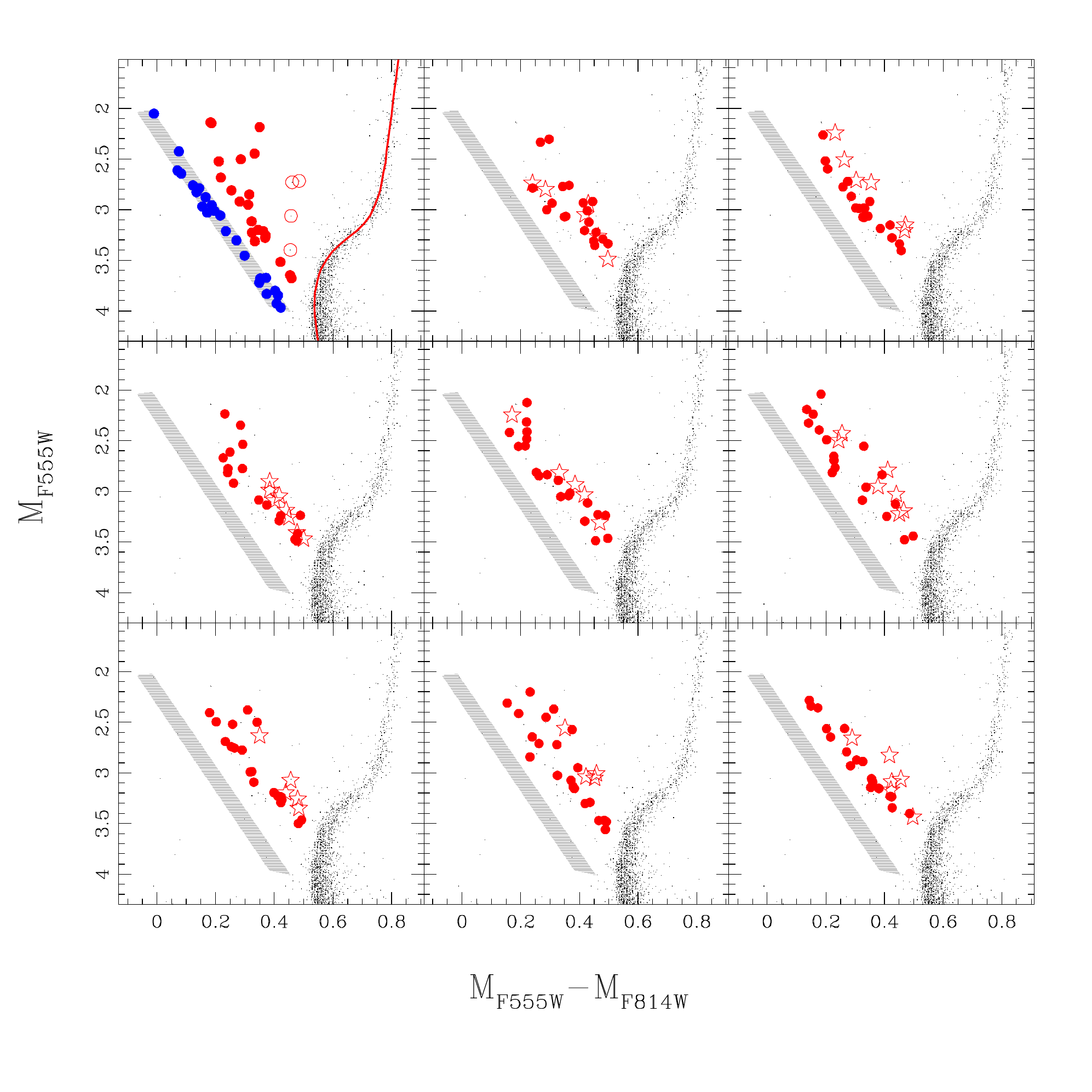}
\caption{Comparison between the synthetic MT-BSS population and
    M30 observations, in the absolute CMD. The upper-left panel shows
    the observations, with all the symbols and the line keeping the
    same meaning as those in Figure 4. The following 8 panels show the
    distribution of 8 sub-samples of 25 synthetic MT-BSSs, with the
    red filled circles representing ``MT-produced'' BSSs, and the red open
    stars corresponding to ``merger-important'' BSSs. In all the
    panels, the grey strip shows the position of the observed blue-BSS
    sequence. }
\end{figure*}

For a proper comparison with the observations, we took into
  account only synthetic binaries that, at the age of 13~Gyr, show
  photometric properties consistent with the actual definition of BSS
  in M30, i.e., that appear bluer and brighter than the current
  location of the cluster MS turnoff point in the CMD.  For each
  Monte-Carlo simulation we obtain on average $\sim 15$ BSSs, the
  maximum number being $\sim 44$.  By performing 2500 such
  simulations, we thus generated a ``reservoir'' of more than
  $\sim4\times10^4$ synthetic BSSs. 
 
Figure 4 shows the distribution of the entire synthetic BSS
sample in the CMD of M30.  The observed CMD (F09) is transferred into
the absolute plane by adopting a distance modulus $(m-M)_V=14.80$ mag
and a color excess $E(V-I)=0.055$ mag (Ferraro et al. 1999b).
Clearly, the distribution of synthetic BSSs (grey dots) well
corresponds to the region where the red-BSSs of M30 (filled red
circles) are observed.  According to the evolutionary tracks shown in
Figure 2, the synthetic MT-BSSs cover a wide strip in the CMD
corresponding to various stages of MT activity during the BSS
formation process.  For this reason, in this work we added to the
observed sample four red-BSSs (empty red circles in Figure 4) that
were not considered in F09, but turned out to be located in the
synthetic MT-BSS region, instead.

Actually, the distribution of the synthetic BSSs shown in Figure 4 can
be used to define a sort of ``MT-BSS domain'' in the CMD, which can be
adopted as reference for future studies.\footnote{\footnotesize
  Nicely, this region well corresponds to the "MT-BSS domain"
  empirically defined by Dalessandro et al. (2013) in the case of NGC
  362.} While a further extension to the red side (i.e., toward lower
surface temperatures) is expected due to the post-MS evolution of
BSSs, the distribution shows a well defined {\it low-luminosity
  boundary} (dashed line in Figure 4), similar to that found in Tian
et al.  (2006).  According to F09, this boundary well corresponds to
the CMD location of the $Z=0.0001$ ZAMS shifted by 0.75 in
magnitude. Very interestingly, this low-luminosity boundary follows
the red side of the gap between the two observed BSS
sequences\footnote{\footnotesize Only five red-BSSs turned out to
    be located (slightly) below the MT low luminosity boundary. This
    is probably due to photometric uncertainties (as some residual
    color equation), or stellar variability. Interestingly, at least
    two of these stars have been classified as W Uma variables by
    Pietrukowicz \& Kaluzny (2004).}, thus demonstrating that no
  MT-BSSs can reach the location of the blue-BSS sequence. In turn,
  this provides further support to the fact that the two BSS sequences
  found in M30 are generated by two distinct formation mechanisms.

The overall distribution of synthetic BSSs shown in Figure 4
  provides a nice view of the MT-BSS domain in the CMD of an old star
  cluster. As the next step, in order to have a more direct comparison
  with the observations (where 25 BSSs are counted along the red
sequence), we performed random extractions of sub-samples of 25
objects from the synthetic MT-BSS reservoir and studied their
distribution in the CMD. To evaluate the relative contribution of MT
and binary merging to the total population, in this analysis we also
distinguished MT-produced BSSs from merger-important BSSs. The results
obtained for 8 random extractions are shown in Figure 5. For the sake
of comparison, the upper-left panel shows the observations, and the
symbols in this panel keep the same meaning as those in Figure 4.  The
populations of 25 synthetic BSSs are presented in the other 8 panels,
where the red solid circles represent MT-produced BSSs, and the red
open stars are merger-important BSSs.  A grey strip corresponding to
the observed blue sequence is also marked for reference in all the
panels. As can be seen from Figure~5, the MT-produced BSSs
  dominate the red population, providing at least $60\%$ of the total
  observed number in all cases.

\section{Summary and discussion}
 M30 is the first star cluster where two distinct sequences of BSSs
 have been observed and have been interpreted as the result of the two
 BSS formation channels (F09): blue-BSSs are generated by stellar
 collisions, red-BSSs derive from MT activity in close
 binaries. Indeed the blue sequence was found to be nicely reproduced
 by collisional isochrones (Sills et al. 2009), while only a
 preliminary guess about the MT-origin for the red-BSSs was provided
 in F09 on the basis of binary evolution models calculated by Tian et
 al. (2006) for the open cluster M67. In this work we presented binary
 evolution models specifically computed for M30 and finally provided
 convincing evidence that the red-sequence is indeed populated by
 MT-BSSs.

We calculated a grid of binary evolution models covering the parameter
space (in terms of masses of the components and orbital separation)
appropriate for the BSSs currently observed in M30 (age=13~Gyr and
$Z=0.0001$). We used Monte-Carlo simulations to randomly generate
large numbers of binary systems, we extracted those binaries that 
can be covered by the grid, we got their physical and
photometric properties at 13~Gyr by interpolating within the model grid, 
and finally we obtained a ``reservoir'' of synthetic BSSs by taking into account 
all the grid-covered MT-binaries that are bluer and brighter
than the MS turnoff of M30. 
The distribution of these objects in the CMD has been
compared to the observed location of the blue- and red-BSS sequences.
Random extractions of 25 such BSSs from the overall reservoir have
been used to investigate the relative importance of MT and
merger processes for the formation of BSSs.  The main results can be
summarized as follows:
\begin{itemize}
\item[1)]The distribution in the CMD of the synthetic MT-BSSs is
  consistent with the location of the observed red sequence
  in M30 and never reaches the region occupied by the blue-BSSs. This
  evidence demonstrates that MT processes are unable to produce such
  ``blue" objects. The result also supports the suggestion (F09) that
  the two parallel sequences observed in M30 are indeed formed by BSSs
  generated by the two distinct formation channels.
\item[2)] Random extractions of 25 synthetic binary-BSSs (the same
  number of observed red-BSSs) show that the models always nicely
  reproduce the observed red sequence.
\item[3)] MT-produced BSSs contribute to at least $60\%$ of the total
  sample, while the remaining $<40\%$ BSSs may require assistance from
  binary mergers. 
\end{itemize}

 Of course, the BSS formation mechanisms are far more complex than
 those investigated in the present paper.  In fact, the internal
 dynamical evolution of GCs certainly plays a significant role in
 mixing BSSs generated by the different channels, especially in the
 cluster cores. Hence, distinguishing the two populations is not an
 easy task. In M30 and a few additional clusters (NGC~362 and
   NGC 1261; see Dalessandro et al. 2013 and Simunovic et al. 2014,
   respectively), two populations of BSSs appear separated by a gap
 in the CMD, thus opening the possibility to investigate them in more
 details. Indeed the evidence collected to date and the work presented
 here suggest that this is a very promising way to distinguish
 collisional from MT BSSs.  However, as pointed out in F09, the GCs in
 which the BSS populations are distinguishable can be rare, since the
 gap separating the two sub-populations is not a permanent feature. In
 fact, the future evolution of the BSSs currently observed along the
 blue-sequence will move these stars toward the red in the CMD, thus
 cleaning up the gap in a few Gyr.  For this reason, a recent burst of
 collisions (possibly driven by the collapse of the core) has been
 suggested to be at the origin of the tiny and well distinct
 blue-sequence observed in M30 (F09).  Thoughtful N-body calculations,
 including internal cluster dynamical evolution, are needed to further
 investigate the properties of the BSS populations in GCs and better
 clarify the physical processes that play the most relevant roles in
 shaping their observed characteristics.

\acknowledgments YX is grateful to Dr. Stephen Justham for the
suggestive discussions. YX also thanks the National Natural Science
Foundation of China for its support through grant Y111221001, and
thanks the 973 Program 2014CB845702. This research is part of the
project {\it Cosmic-Lab} (web site: http://www.cosmic-lab.eu) funded
by the European Research Council (under contract ERC-2010-AdG-267675).
We thank the anonymous referee for useful suggestions.

\clearpage

\end{document}